\documentclass{amsart}
\usepackage{amsmath}
\usepackage{amsfonts}
\usepackage{graphics}
\usepackage{epsfig}
\usepackage{amssymb}
\usepackage{amscd}
\usepackage[all]{xy}
\usepackage{latexsym}
\usepackage{graphicx}
\usepackage{multirow}
\usepackage{geometry}

\theoremstyle{remark}

\newtheorem{thm}{\bf Theorem}[section]
\newtheorem{cor}{\bf Corollary}[section]

\title{Quantum curves}
\author{Albert Schwarz}
\address{Department of Mathematics\\
University of California\\
Davis, CA 95616, USA\\ schwarz@math.ucdavis.edu}

\begin{document}
\maketitle
{\bf Abstract}

One says  that a pair $(P,Q)$ of ordinary differential operators specify a quantum curve if $[P,Q]=\hbar$. If a pair of difference operators $(K,L)$ obey the relation $KL=q LK$ where $q =e^{\hbar}$  we say that they specify a discrete quantum curve.

This terminology is prompted by well known results about commuting differential and difference operators , relating pairs of such operators with pairs of meromorphic functions on algebraic curves obeying some conditions.

The goal of this paper is to study the moduli spaces of quantum curves. We will relate the moduli spaces for different $\hbar$. We will show how to quantize a pair of commuting differential or difference operators (i.e. to construct the corresponding quantum curve or discrete quantum curve)

\section{Introduction}
%\subsection{}
One says  that a pair $(P,Q)$ of ordinary differential operators specify a quantum curve if $[P,Q]=\hbar$ \cite {M},\cite {SCH},\cite {D}. If a pair of difference operators $(K,L)$ obey the relation $KL=\lambda LK$ where $\lambda =e^{\hbar}$  we say that they specify a discrete quantum curve (we can impose also an additional condition that $L^{-1}$ is a difference operator in the definition of quantum curve).

This terminology is prompted by well known results about commuting differential and difference operators \cite {KR},\cite {MUM}, relating pairs of such operators with pairs of meromorphic functions on algebraic curves obeying some conditions.

The goal of this paper is to study the moduli spaces of quantum curves. We will relate the moduli spaces for different $\hbar$. We will show how to quantize a pair of commuting differential or difference operators (i.e. to construct the corresponding quantum curve or discrete quantum curve). This construction generalizes the considerations of \cite {SCH}.

The KP-hierarchy acts on the moduli space of quantum curves; we prove that similarly the discrete KP-hierarchy acts on the moduli space of discrete quantum curves.

We consider  also matrix differential and difference operators and obtain  similar results. (The generalization of \cite {SCH} to matrix differential operators was given in \cite {KV}.)

Eynard-Orantin topological recursion \cite {EO} gives a construction of free energy and correlation functions corresponding to an algebraic curves and two meromorphic functions on it.  We construct a
quantum curve starting with the same data (but  the conditions that we impose on meromorphic functions are different).  It seems that the paper \cite {GS} can be considered as a bridge between topological recursion and our constructions.  Another way to relate topological recursion to quantum curves in our sense can be based on the comparison of Virasoro constraints that can be derived in both situations. The modification of topological recursion that was required to "remodel B-model" \cite {EM}  should be related to discrete quantum curves in our sense. 

We will discuss the relation of our constructions to the results of \cite {D}.

\section{Differential operators. Quantum curves.}
Let us define a pseudodifferential operator as a formal series 
$$L=\sum a_k(x) D^k$$
where $D=\frac{d}{dx}$ and $a_k(x)$  stands for a formal power series:  $a_k(x)=\sum a_{kl}x^l.$  We assume that $k\in \mathbb{Z}$ and $a_k(x)=0$ for $k>>0$.    The operator has order  $q$ if  its leading term (the non-zero term with greatest $k$) is equal to $a_q(x)D^q$; the operator is monic if $a_q(x)=1$, a monic operator is normalized if $a_{q-1}(x)=0.$ Monic pseudodifferential operators of order $0$ form a group denoted by $\mathcal{G}.$  \footnote {Pseudodifferential operators constitute an associative algebra $\mathcal{A}$ for appropriate definition of multiplication. We will not give this definition; see, for example, \cite {MUL}. Let us notice, however, that the multiplicative structure in $\mathcal{A}$ can be recovered from the representation of $\mathcal{A}$ by the operators in $\mathcal{H}$ that is described in the next paragraph.}

We denote by $\mathcal{H}$ the space of Laurent series $\sum _{k<<\infty}c_kz^k$ ,  by
$\mathcal{H}_+$ its subspace consisting of polynomials and by $\mathcal{H}_-$ the subspace  of Laurent series obeying $c_n=0$ for $n\geq 0 .$ Pseudodifferential operators act on $\mathcal{H}$; the differentiation $D$ acts as multiplication by $z$ and multiplication by $x$ acts  as $-\frac{d}{dz}.$ Differential operators can be characterized  as  pseudodifferential operators preserving $\mathcal{H}_+.$ Every pseudodifferential operator $L$ can be represented as a sum of differential operator 
$L_+=\sum _{k\geq 0}a_k(x) D^k$ and "integral" operator $L_-=\sum_{k<0} a_k(x) D^k.$

We will denote   by $Gr$ the space of all subspaces $V\subset \mathcal{H}$ such that the natural projection $\pi_+: \mathcal{H}\to\mathcal{H}_+$ induces an isomorphism between $V$ and $\mathcal{H}_+.$ In other words the subspace $V\in Gr$ if it has a basis of the form $v_n=z^n+r_n$ where $n\geq 0$ and $r_n\in \mathcal{H}_-.$ The space $Gr$ is called Sato Grassmannian. \footnote {It is the big cell of  the index zero part  of the infinite-dimensional  Grassmannian; we do not need the description of  this  Grassmannian (see, for example, \cite {MUL}).}

 The following theorems belong to Sato (see \cite {MUL} for the proof):
 \begin{thm}
 \label{1}
 There exists one-to-one correspondence between the elements of  the group $\mathcal{G}$ of monic zeroth  order differential operators and points of $Gr$. Namely, every subspace $V\in Gr$ has a unique representation in the form $V=S\mathcal{H}_+$ where $S\in \mathcal{G}.$
\end{thm}

The commutative Lie algebra $\gamma_+$ of polynomials $\sum_{k\geq 0} t_iz^i$ acts on $Gr$ in natural way. (This action comes from the remark that we can multiply the elements of $\mathcal{H}$ by $g(t)=\exp (\sum_{k\geq 0} t_iz^i) $  where $t_i$ are nilpotent parameters.) \footnote {More generally, one can consider the action of the Lie algebra $\gamma$ of polynomials  $\sum_{-\infty<<k <<\infty} t_iz^i$; this action will be important in the next section.}  

It is clear  from Theorem \ref {1} that $\gamma_+$  acfs also on $\mathcal{G:}$ 
\begin{equation}
\label{s}
\frac{\partial S}{\partial t_n}=(SD^nS^{-1})_-S.
\end{equation}
\begin {thm}
\label{11}
Every normalized pseudodifferential operator $Q$ of order $q$ can be represented in the form $ S^{-1}D^qS$ where $S\in \mathcal{G}$ ; this representation is unique up to multiplication by an operator with constant coefficients.
\end{thm} 
Using this statement  we can construct the action of Lie algebra $\gamma_+$  on the space of normalized pseudodifferential operators  of order $q$ differentiating the relation $Q(t)= S^{-1}(t)D^qS(t)$ with respect  to $t_n.$

The action on this space can be written in the form of differential equation
\begin{equation}
\label{kp}
\frac{\partial Q}{\partial t_n}=[Q^{\frac{n}{q}}_+,Q]
\end{equation}
Notice that  this formula determines also the action of Lie algebra  $\gamma_+$ on the space of normalized differential operators.  {\footnote { The fractional powers entering this formula can be defined using the representation $Q=S^{-1}D^qS$.}}

All actions we described  can be considered as different forms of KP-hierarchy.

We would like to solve the equation $[P,Q]=\hbar.$ We assume that $P$ is a differential operator  of order $p$ and $Q$ is a normalized differential operator of order $q$.  Using  Theorem \ref {11} we construct the operator $S\in \mathcal{G}$  such that $SQS^{-1}=D^q$ . Introducing the notation
$V=S\mathcal{H}_+$ we obtain a subspace $V\in Gr$ invariant with respect to multiplication by $z^q$ and with respect to the action of the operator $\tilde P= \hbar \frac{d}{dz^q}+b(z)$ where $b(z)$ stands for the multiplication by a Laurent series denoted by the same letter. (We use the fact that the action of $D^q$ can be interpreted as multiplication by $z^q$ and the fact that $\mathcal{H}_+$ is invariant with respect to the action of differential operators. The form of the operator $\tilde P=SPS^{-1}$ follows from the relation $[\tilde P,\tilde Q]=\hbar$ where $\tilde Q=SQS^{-1}=z^q.$)   We can invert this  consideration to obtain the following statement \cite {SCH}:
\begin{thm} \label {2}
If $V\in Gr$ is invariant  with respect to the operator of multiplication by $z^q$ and with respect  to the operator $\hbar \frac{d}{dz^q}+b(z)$ we can construct  a  differential operator $P$ and a normalized differential operator $Q$ obeying $[P,Q]=\hbar.$ The leading term of the operator $P$ is determined by the leading term of the Laurent series $b(z).$
\end{thm}
The construction is based on the Sato theorem: We represent  $V$ in the form $V=S\mathcal{H}_+$ where $S\in \mathcal{G}$ and transform  the operators acting on $V$ by means of the operator $S$. We obtain pseudodifferential operators acting in $\mathcal{H}_+$, i.e. differential operators.

Notice that the points of Grassmannian that are invariant with respect to  multiplication by $z^q$ and with respect to the action of the operator of the form $\hbar \frac{d}{dz^q}+b(z)$
appeared for first time in the study of partition function of 2D-gravity  in the paper \cite {KS} ; they were studied later in numerous papers (sometimes these points are called string points of the Grassmannian; the operators preserving them are called Kac-Schwarz operators). A point  $V\in Gr$ that is  invariant with respect  to multiplication by $z^q$ and with respect to the action of the operator  $A=\hbar \frac{d}{dz^q}+b(z)$ is invariant also with respect to the operators $z^{nq}A$ where $q=0,1,2, \dots$; it was shown in \cite {KS} that this implies Virasoro constraints on the  state corresponding to $V$. (Recall that for every point $V\in Gr$  one can construct a state $\Psi _V$ in fermionic Fock space and  a bosonic state represented by tau-function. If  $V$ is invariant with respect to the operator $C$ having matrix elements $c_{mn}$ in the basis $z^n\in 
\mathcal{H})$ then $\Psi _V$ is an eigenvector of the operator $\sum c_{mn}:\psi_m\psi_n^+:$ where $\psi _m$ are operators obeying canonical anticommutation relations; see \cite {KS}. This remark allows us to derive the Virasoro constraints for fermionic state and for tau-function.)

 It follows from Theorem \ref {2}    that the  Lie algebra $\gamma _+$ acts  on the space of pairs $(P,Q)$ of differential operators obeying    $[P,Q]=\hbar.$ (We assume that $Q$ is normalized.) The proof is based on the remark that for $V\in Gr$ satisfying the conditions of the theorem the subspace $V(t)=g(t)V$ also obeys the same conditions with $\tilde P=\hbar \frac{d}{dz^q}+b(z)$ replaced with 
$$g^{-1}(t)\tilde P g(t)=\hbar \frac{d}{dz^q}+b(z)-\sum \frac{i}{q}t_iz^{i-q}.$$

In other words, we can say that the KP-flows (\ref {kp}) are defined on the space of pairs we are interested in.

We say that the vectors $v_0, ....,v_{q-1}$ form a $g$-basis in $V$ if the vectors $g^mv_i$
where $0\leq i<q,  0\leq m$,  form a basis of $V.$ We will  use this definition in the case $g=z^q.$

To construct an example of $z^q$ - basis we recall that $V\in Gr$ has a basis $v_n=z^n+r_n$ where $r_n\in \mathcal{H}_-.$ The first $q$ vectors 
of this basis (vectors $v_0, ....,v_{q-1}$ ) form a $z^q$-basis of $V.$ 

If  $SQS^{-1}=D^q$ , $V=S\mathcal{H}_+$ , $v_0, ....,v_{q-1}$ is a $z^q$-basis of $V$ we represent the operator  $\tilde P=SPS^{-1}$ in this basis.
We obtain
\begin{equation}
\label{ }
\tilde Pv_i=M^j_i(z^q)v_j
\end{equation}
where the entries of the matrix $M$ are polynomials with respect to $z^q$.  We say that the matrix $M$ is the companion matrix of the pair $(P,Q)$. (Alternatively one can define the companion matrix as a matrix of $P$ in a $Q$-basis of $\mathcal{H}_+.$)  The companion matrix depends on the choice of  $z^q$-basis.  Starting with a  $z^q$-basis $v_i$  we can construct a new $z^q$-basis by  the formula $ v' _i=g_i^j(z^q)v_j$ where $g_i^j(z^q)$ is an invertible polynomial matrix (its entries   are polynomials with respect to $z^q$). The transformation rule for the matrix $M$ by the change of $z^q$-basis has the form of gauge transformation : $ M'=gMg^{-1}+\hbar \frac{dg}{dz^q}g^{-1}.$ This means that $M$ specifies a connection.  {\it {We will  choose the basis $v_i$ in such a way that 
$v_i=c_iz^i+$lower order terms}}, $c_i\neq 0$. (Notice that this condition does not specify the vectors $v_i$ uniquely; we can replace $v_i$ with $\tilde v_i=t_i^jv_j$ where $ t$ is a constant invertible triangular matrix: $t_i^j=0$  for $i\leq j$.)

 Let us prove the following theorem:
\begin{thm}\label{3}
If the leading coefficient (the coefficient of the leading term)  of the matrix $z^{j-i}M^j_i(z^q)$ has $q$ distinct  eigenvalues then there exist $q$ pairs     of differential operators obeying $[P,Q]=\hbar$ and having $M$ as the companion matrix.
\end{thm}
To prove this theorem we should find $q$ Laurent series $b(z)$ and corresponding $v_0, ....,v_{q-1}$ in such a way that
\begin{equation}
\label{4}
(\hbar \frac{d}{dz^q}+b(z))v_i=M^j_i(z^q)v_j
\end{equation}
and $z^{mq}v_i$
where $0\leq i<q, 0\leq m$ form a basis of a subspace $V\in Gr.$ 
Then  we can apply Theorem \ref {2} .

We  change the variables in the equation (\ref{4}) substituting $v_i=z^iu_i$ where $u_i=c_i+$lower order terms.  We obtain the equation
\begin{equation}
\label{5}
(\hbar \frac{d}{dz^q}+b(z))u_i=  B^j_i(z)u _j
\end{equation}
where
$$B(z)=(B_i^j(z)) = \left(M_i^j(z^q) z^{j-i}-\frac{i\hbar}{q z^q}\delta_i^j\right)$$

Let us consider first the case when $\hbar=0.$ Then $b(z)$ is one of  eigenvalues $\lambda _k(z)$ of the matrix $B^j_i(z)$ and $u_i$ are components of the eigenvector. The existence of $b$ and of the vector $u(z)=(u_0,...,u_{q-1} ) $ obeying the conditions we need follows immediately from the perturbation theory . (If $B$ is replaced by its leading term this statement follows from our assumptions. All other terms of $B$ can be considered as a perturbation of the leading term. The leading term  of the matrix $ B^j_i(z)$ coincides with the leading term of the matrix $z^{j-i}M^j_i(z^q),$  therefore its eigenvalues are distinct. This allows us to construct $b$ as a Laurent series and $u(z)$ as a power series with respect to $z^{-1}$.)

If $\hbar\neq 0$  we consider the auxiliary equation
\begin{equation}
\label{6}
\hbar \frac{d}{dz^q}w_i=  B^j_i(z)w _j.
\end{equation}
or, equivalently,
\begin{equation}
\label{7}
\hbar \frac{d}{dz}w_i= qz^{q-1} B^j_i(z)w _j.
\end{equation}

As we have noticed the eigenvalues of the leading term of $B(z)$ are distinct.
 This allows us to say that the equation can be diagonalized by means of the formal change of variables $w(z)=R(z)t(z)$ where $R(z)=1+\sum _{k\geq 1}R_kz^{-k}$ ; see \cite {WAS}. This means that the equation for the components of the vector $t(z)$ looks as follows:
\begin{equation}
\label{8}
\hbar \frac{d}{dz^q}t_i=\Lambda_i t_i.
\end{equation}

Let  us consider $q$ solutions of the equation (\ref{8}) having the form
$$t_k=\exp(\int {\hbar}^{-1}\Lambda_k(z)qz^{q-1}dz),$$
$t_i=0$ for  $i\neq k.$

The corresponding solutions of the equation (\ref {6})   have the form
$$ w_i=\exp(\int {\hbar}^{-1}\Lambda_k(z)qz^{q-1}dz)r_{ik}(z),$$
where $r_{ik}(z)= c_{ik}+$ lower order terms.
Now for every $k$ it is easy to find $b(z)$ in such a way that $r_{ik}(z)$ becomes a solution to the equation (\ref{5}). Namely, we should take
$$b_k(z)=\Lambda_k(z).$$
(We use the fact that the equation (\ref {5}) can be reduced to the equation (\ref{6}) by means of the substitution $w= \rho u.$) This gives the proof of the theorem.

A  shorter proof can be given in the following way. Notice that the formal change of variables $w(z)=R(z)t(z)$  transforms (\ref {6}) into equation
\begin{equation}
\label{88}
\hbar \frac{d}{dz^q}t_i=\Lambda_i^j t_j.
\end{equation}
where
\begin {equation}
\label {89}
\Lambda_i ^j(z)=
S_i^rB_r^mR^j_m-\hbar S_i^m\frac{dR_m^J}{dz^q},
\end{equation}
and $S$ denotes the matrix inverse to the matrix $R.$

We choose $R$ in such a way that the matrix $\Lambda$ is a diagonal matrix with entries $\Lambda _i.$ Then it follows from (\ref {89}) that for every $r$ the series $u_i(z)=R_i^r(z)$ 
is a solution of (\ref {5}) with $b(z)=\Lambda _r(z).$

Notice that 
$\Lambda _k(z)=\lambda _k(z)+O(\hbar)$
where $\lambda_k(z)$ stands for  the eigenvalue of the matrix $B^j_i(z)$. (This follows, for example, from the comparison with the case $\hbar=0$).

Let us give another proof of the theorem that can be applied in more general situations.
We would like to find solutions of the equation (\ref {5}) as power series :
$$u(z)=\sum _{k\geq 0} {^ku} z^{-k},$$
$$b(z)=\sum _{k\geq 0} { ^kb }z^{p-k}.$$
We introduce the notation
$$B=\sum _{k\geq 0}{^kB} z^{p-k}.$$
Here $u(z)$ and the corresponding coefficients $^k u$ are  considered as $k$-dimensional vectors; $B(z)$ and $^kB$ are  $q\times q$ dimensional matrices. We can solve the equation (\ref {5}). The recursion formula looks as follows:
\begin{equation}
\label{8}
({^0b}-{^0B})(^ku)=({^kB}-{^kb})(^0u) +\rm {known}\: \rm {terms}.
\end{equation}

In particular,
$$({^0b}-{^0B})({^0u})=0,$$
i.e. $^0b$ is an eigenvalue of $^0B.$ 
 It follows from our assumptions that all eigenvalues of $^0B$ (that coincides with the leading coefficient of  $z^{j-i}M^j_i(z^q)$) are simple. This means that the image of the operator $^0b-^0B$
 has codimension $1$. We denote by $\rho $  a non-zero linear functional vanishing on this image. (It can be interpreted as an eigenvector the matrix transposed to $^0B$ with eigenvalue $^0b$.)   Applying $\rho$ to both parts of recursion formula and  noticing that $\rho(^0u)\neq 0$ we can calculate $^kb$. Then the recursion formula gives us $^ku.$ Noticing that we can take any eigenvalue of $^0B$ as $^0b$ we obtain the proof of the theorem. %(The ambiguity in the calculation of $^ku$ is irrelevant. It  is related to the freedom in the choice of $z^q$ basis.??)

The group $C_q$ of $q$-th roots of unity acts on solutions of the equation (\ref{6}) (if ($u_0(z),...,u_{q-1}(z))$ is a solution and $\epsilon^q=1$ then $(u_0(\epsilon z),...., \epsilon^{-i}u_i(\epsilon z),...)$ is again a solution). One can use this fact to check that the group $C_q$ acts also on the set of solutions constructed above. If $p$ and $q$ are coprime then this action is transitive therefore for appropriate labeling $\Lambda_{k+1}(z)=\Lambda_k (\epsilon z) .$ (Here $\epsilon $ stands for  a  primitive root of unity.) 
Similarly,  we can assume that $\lambda_{k+1}(z)=\lambda_k (\epsilon z) .$ From this equation one can derive the properties of eigenvalues of the leading term of the matrix   $B^j_i(z)$. If this  leading term  has degree $p$ then the leading terms of eigenvalues have the same degree:
$\lambda_k(z)=\alpha_kz^p+...$ and we obtain that $\alpha_{k+1}=\epsilon ^{p}\alpha_k$, hence
$\alpha _k=\epsilon ^{kp}\alpha_1$. Therefore in the case when $p$ and $q$ are coprime the numbers $\alpha_k$ are distinct.

Let us consider  pairs $(P,Q)$ where $P$ is a  differential operator of order  $p$, $Q$ is a normalized differential operator of order $q$, the orders $p$ and $q$ are coprime and $[P,Q]=\hbar$.     It follows from the above considerations that the order of $b(z)$ is equal to $p$, therefore the leading coefficient of the matrix  $B^j_i(z)$ (coinciding with  the leading coefficient  of the matrix $z^{j-i}M^j_i(z^q))$ has distinct eigenvalues. This allows us to describe the moduli space of such pairs and to prove that this moduli space does not depend on $\hbar$ (see \cite {SCH}). We can formulate this description in the following way.

Let us say that the polynomial $q\times q$ matrix $M^j_i(z^q)$ is regular if the leading coefficient  of the matrix $z^{j-i}M^j_i(z^q))$ has $q$ distinct eigenvalues. If this leading coefficient has degree $p$ we say that a regular matrix $M$  belongs to the space $\mathcal{M}_{p,q}.$
We say that a solution of the equation $[P,Q]=\hbar$ where $P$ is a  differential operator of order  $p$, $Q$ is a normalized differential operator of order $q$ is regular if   the companion matrix $M$ is regular.  It follows 
from the above consideration that the moduli space of regular solutions does not depend on $\hbar.$ It can be considered as a $q$-fold covering of the space  $\mathcal{M}_{p,q}/\mathcal{T}$ where $\mathcal{T}$ denotes the group of triangular matrices. If $p$ and $q$ are coprime all solutions of the equation $[P,Q]=\hbar$ are regular.

In the above statements we  have used the choice of $z^q$-basis specified by the condition $v_i=c_iz^i+$lower order terms.  We say that a companion matrix in general $z^q$-basis is regular, if the corresponding matrix in the preferred basis is regular. The condition of regularity for other choices of $z^q$-basis is not so simple. However, one can give a necessary condition of regularity that is valid in any basis. Namely, we should consider the characteristic polynomial
$$\det(M-\lambda\cdot 1)=\sum A_i(z^q)\lambda^i.$$
Let us suppose that $\det M=A_0(z^q)$ is a polynomial of degree $p$ with respect to $z^q$. If $M$ is regular then the degree of $A_i$ with respect to $z^q$ is less or equal than $[\frac{p(q-i)}{q}].$ This follows from the remark that the asymptotic behavior of the characteristic polynomial of matrix $M$  for large $z$ does not depend on the choice of basis.

The results we have obtained for scalar differential operators can be generalized to the case of matrix differential operators. Instead of  Grassmannian $Gr$ we should consider the vector Sato Grassmannian $Gr_s$.  It consists of subspaces $V$ of $\mathcal {H}^s =\mathcal{H}\otimes \mathbb{C}^s$ (of direct sum of $s$ copies of $\mathcal{H}$) such that the natural projection  of $V$ to  $\mathcal {H}_+^s=\mathcal{H}_+\otimes \mathbb{C}^s $ is an isomorphism.

It is easy to generalize Theorem \ref {1} and Theorem \ref {2} to this case; see  \cite{LiMUL}, Th.6.2, \cite {KV}, Prop 2.1. 
 \begin{thm} \label {g}
If $V\in Gr_s$ is invariant  with respect to the operator of multiplication by $z^q$ and with respect  to the operator $\tilde P=\hbar \frac{d}{dz^q}+b(z)$ we can construct  a  matrix differential operator $P$ and a normalized matrix differential operator $Q$ of  order $q$ obeying $[P,Q]=\hbar.$  Here $b(z)$ is a Laurent series having $s\times s$ matrices as coefficients; if its leading term has degree $p$ then the operator $P$ is of order $p.$ 
\end{thm}

Notice that as in the scalar case one can  derive the Virasoro constraints on the fermionic state $\Psi _V$  and on the corresponding tau-function.

Again we can define the companion matrix of the pair $(P,Q)$ as a matrix of $\tilde P$ in a $z^q$-basis of $V.$ It is convenient to use a $z^q$-basis obeying 

$v_{i\alpha}=c_{i\alpha}z^ie_{\alpha}$+lower order terms.

Here $0\leq i<q$,$1\leq \alpha\leq s$, $e_{\alpha}$ denotes the standard basis of  $\mathbb{C}^s$
and $c_{i\alpha}$ are non-vanishing constants.

The companion matrix  in this basis can be regarded as a $q\times q$ matrix with  with entries that  are $s\times s$ matrices depending polynomially on $z^q$. We denote this matrix as  $M_i^j.$  It is defined by the equation  (\ref {4}) where $v_i$ denotes now a $q$-dimensional vector having $s$-dimensional vectors as components. Introducing  the notation  $u _{i\alpha}=z^{-i}v_{i \alpha}$ we obtain the equation (\ref {5}) where the matrix $B$ is defined by the same formula as in scalar case.

We will prove the following theorem:
\begin{thm}\label{3m}
Let us suppose that the entries of  the leading coefficient  of the matrix $z^{j-i}M^j_i(z^q)$ are scalar  matrices. In other words we assume that $^0B$ (the leading  coefficient of the matrix  $B$) has the form $^0B=\sigma\otimes I_s$ where $\sigma$ is  a $q\times q$ matrix with complex entries and $I_s$ stands for unit $s\times s$ matrix; we assume that $\sigma$ has $q$ distinct  eigenvalues. Then there exist $q$ pairs     of  matrix  differential operators obeying $[P,Q]=\hbar$ and having $M$ as the companion matrix.
\end{thm}
We derive this statement from Theorem \ref {g} generalizing the second proof of Theorem \ref {3}. It is sufficient to check that the equation (\ref {5}) has $q$ solutions obeying

$u_{i\alpha}=c_{i\alpha} +$lower order terms. 

Then we can define $V$ as a subspace spanned by $z^{qm}v_{i\alpha}$ where $v_{i\alpha}=z^iu_{i\alpha}.$

The recursion formula (\ref {8}) for the solution of (\ref{5} ) can be written in more detail in the form

$$({^0b_{\alpha}^{\beta}}\delta _i^j-{^0B_{i\alpha}^{j\beta}})(^ku_{j\beta})=({^kB_{i\alpha}^{j\beta}}-{^kb_{\alpha}^{\beta}}
\delta _i^j)(^0u_{j\beta}) +\rm {known}\: \rm {terms}.$$

In particular,
$$({^0b_{\alpha}^{\beta}}\delta _i^j-{^0B}_{i\alpha}^{j\beta})(^0u_{j\beta})=0.$$

Recall that ${^0B}_{i\alpha}^{j\beta}=\sigma_i^j\delta _{\alpha}^{\beta}$; if $\sigma_i^j s_j=\lambda s_i $ (i.e. $s$ is an eigenvector of $ \sigma$ ) we can take
${^0u}_{i\alpha}=s_i$.  Let us denote by $\rho$ the eigenvector of the transposed matrix $\sigma$ having eigenvalue $\lambda$; it follows from our assumptions that $<\rho,s>\neq 0.$ The inner  product  of LHS of the recursion relation with $\rho$ vanishes; this allows us to calculate ${^kb_{\alpha}^{\beta}}.$
Then the recursion formula gives us ${^ku}_{i\alpha}.$

We have assumed in Theorem \ref {3m} that $\sigma$ has $q$ distinct eigenvalues.  This condition is satisfied if the degree $p$ of the leading term of $B$ is coprime with $q$; the proof is the same as in scalar case.
 \section {Difference operators.  Discrete quantum curves.}
 
 We would like to solve the equation $KL=qLK$ where $K$ and $L$ are difference operators.  Our consideration will be based on the notion of pseudodifference operator.
 
 Let us consider  linear operators acting on the space   of Laurent series $\mathcal{H}$ . We say that a (doubly infinite) sequence $a=(a_k)$ specifies a diagonal operator transforming a sequence $c_k$ into a sequence $a_kc_k$ (or equivalently a series $\sum _{-\infty <k <<\infty}c_kz^k$ into the series  $\sum _{-\infty <k <<\infty}a_kc_kz^k$); we denote this operator by the same letter $a.$ The shift operator $\Lambda$ transforms a sequence $c_k$ into the sequence $c_{k-1}$ (equivalently  the series  $c(z)=\sum _{-\infty <k <<\infty} c_kz^k$  goes to the series $zc(z)= \sum _{-\infty <k <<\infty}c_{k-1}z^k)).$ In other words $\Lambda $ can be interpreted as multiplication by $z.$
 
 Notice, that  one can  consider $k$ as a continuous parameter $k\in \mathbb{R}.$ Then we can modify the definition of the shift operator considering the operator $\Lambda_{\hbar}$ that transforms $c(k)$ into $c(k-\hbar)$. Of course, for fixed $\hbar$ this makes no difference. However, in applications we should  consider $\hbar$ as a small parameter and work with power series with  respect to $\hbar.$
 
 One defines pseudodifference operators by the formula
 $$L=\sum_{-\infty <s <<\infty} a(s) \Lambda ^s$$
 where $a(s)$ are diagonal operators. If the sum is finite then $L$ is a difference operator. Restricting the summation to negative $s$ we obtain the operator $L_-$. Taking the sum over $s\geq 0$ we obtain the operator $L_+$; notice that $L_+$ is a difference operator. 
 
 An operator of the form 
 $$\Lambda ^n+\sum _{-\infty <s<n} a(s) \Lambda ^s$$
 is a monic pseudodifference operator of order $n.$  Monic   pseudodifference operators of order zero form a group denoted by  $\mathcal{S}.$
 
 The space $\mathcal{H}$ has natural decreasing filtration $H_m=z^m\mathcal{H}_+=span (z^m,z^{m+1}, \cdots).$ One can characterize difference operators as pseudodifference operators compatible with this filtration.\footnote { We say that an  operator  $A$ and  a decreasing  filtration $F_m$ are compatible if for some integer $a$ we have  $AF_m\subset F_{m-a}$ for all $n.$} One can give another characterization of difference operators as pseudodifference operators transforming the space of Laurent polynomials $H=\bigcup H_m$ into itself.
 
 Let us say that a flag  in $\mathcal{H}$ is a decreasing filtration $V_m$ such that $z^{-m}V_m\in Gr.$ In other words we assume that the natural projection $\pi_m:V_m\to H_m$ is an isomorphism. Let us denote by $w_m(z)$ the point of $V_m$ obeying $\pi_m(w_m)=z^m$.
 A flag is specified by an arbitrary sequence of series $w_m(z)$ obeying the condition
 $$w_m(z)=z^m+\sum_{k<m} w_{m,k}z^k,$$
 hence $w_{m,k}$ can be considered as coordinates in the space $\mathcal{F}$ of flags.
 (We should take $V_m=span (w_m,w_{m+1},\cdots).$)
 
  The notion of flag was essentially used in \cite {AM}, the theorems below are closely related to the results of this paper. However, they admit simple independent proofs.

\begin {thm}
\label {d}
For every monic pseudodifference operator  of order zero $S\in \mathcal{S}$ we can construct a flag $V$  in $\mathcal{H}$  taking $V_m=SH_m.$ This construction gives a one-to-one correspondence between $\mathcal{S}$ and the space of flags $\mathcal{F}$. 
\end{thm}

To prove this fact we notice that the operator $S=1+\sum_{r>0}s(r)\Lambda^{-r}$ where $s(r)$ are diagonal operators transforms $z^m$ into 
$$S(z^m)=z^m+\sum_{r>0}s_{m-r}(r)z^{m-r}.$$
We see immediately that $SH_m=span (  S(z^m),S(z^{m+1}) ,\cdots)$ is a flag in $\mathcal{H}.$ 
The map $\mathcal{S}\to \mathcal{F}$ is bijective (we can identify the coordinates in these spaces using the formula $s_{m-r}(r)=w_{m, m-r}$).

\begin {cor}
The representation $V_m=SH_m$ induces a correspondence between pseudodifference operators  compatible with the flag $V_m$ and difference operators (=pseudodifference operators compatible with the flag $H_m$).
\end {cor}
We  define the flag space $V$ as the union of spaces $V_m$ ; in particular,  $H$ is a flag space of the flag $H_m$.
 \begin {cor}
 The representation $V_m=SH_m$ induces a correspondence between pseudodifference operators preserving the flag space $V=\bigcup V_m$ and difference operators.
 \end {cor}
\begin {thm}\label {ps}
Every monic pseudodifference operator $L$ of order $n$ can be transformed into operator $\Lambda^n$ by means of monic pseudodifference operator of order zero:
\begin{equation}
\label{s}
\Lambda ^n=SLS^{-1}, 
\end{equation}
 where  ${S\in \mathcal{S}}$. If $L$ is a difference operator then the flag space $V=SH$ is $\Lambda^n$-invariant. If
 $L$ is an invertible difference operator (i.e. $L^{-1}$ exists and also is a difference operator) then $V$ is $\Lambda^{-n}$-invariant. 
\end{thm}
This theorem can be used to give a general construction of invertible difference operators.
\begin {cor}
Let us suppose the flag space $V=\bigcup V_m$ is invariant with respect to multiplication by $z^n$ and by $z^{-n}.$
Then representing $V$ in the form $V=SH$ where $S\in \mathcal{S}$ we can construct an invertible difference operator $L$ using the formula $L=S^{-1}\Lambda^nS.$
\end {cor}
To check this fact we notice that $H$ is invariant with respect to $L$ and $L^{-1}=S^{-1}\Lambda^{-n}S.$

The commutative Lie algebra $\gamma$ of polynomials $\sum_{-\infty<<k <<\infty} t_iz^i$ acts on $\mathcal{H}$, hence on $Gr$ and on  the space of flags $\mathcal{F}. $  As in  Section 2 this follows from the remark that   for $g(t)=\exp (\sum_{k\geq 0} t_iz^i) $  where $t_i$ are nilpotent parameters and a flag $V=(V_m)$ we can consider a flag $g(t)V=(g(t)V_m)$.

We can use Theorem \ref {d} to define the action of the Lie algebra $\gamma$ on  $\mathcal{G}$
and Theorem \ref {ps} to define the action on monic pseudodifference operators of order $n$. These actions can be written in the form
\begin{equation}
\label{dkp }
\frac{\partial S}{\partial t_m}=(S\Lambda ^mS^{-1})_-S
\end{equation}
\begin{equation}
\label{ekp}
\frac{\partial L}{\partial t_m}=[L^{\frac{m}{n}}_+,L]
\end{equation}
These formulas are called discrete KP equations  (or Toda equations) \cite {AM}. They are very similar to formulas of Section 2. 
However, here $m\in \mathbb{Z}$, all operators are pseudodifference operators, $D$ is replaced by $\Lambda.$ Notice, that formula (\ref{ekp}) can be considered also as an action  of $\gamma$ on the space of monic difference operators.

To solve the equation $KL=qLK$ where $L$ is a monic  difference operator we consider the  flag $V_n=SH_n$ where $S$  is defined by the formula (\ref {s}).
If $K$ is a difference operators then the flag $V_m=SH_m$ is compatible with operators 
$\tilde K=SKS^{-1}$ and $\Lambda ^n=SLS^{-1}. $  Using Theorem \ref {d} we obtain
\begin {thm}
\label {e}
For every pair $(K,L)$ of difference operators obeying $KL=qLK$ where $L$ is monic of order $n$  one can construct a  flag $V_m$ in $\mathcal{H}$ such that it is compatible with operator  $\Lambda ^n$ and  with pseudodifference operator $\tilde K$  obeying $\tilde K\Lambda ^n=q \Lambda ^n \tilde K.$

Conversely, if we have  a   flag $V_m$ in $\mathcal{H}$ such that it is compatible with operator  $\Lambda ^n$ and  with pseudodifference operator $\tilde K$ obeying  $\tilde K\Lambda ^n=q \Lambda ^n \tilde K$
 we can construct a pair $(K,L)$ of difference operators obeying $KL=qLK.$ 
\end {thm}

Recall that the operator $\Lambda ^n$ acts as multiplication by $z^n.$ The operator $\tilde K$ can be represented in the form
$$\tilde K=({^0b})\Lambda ^p+ ({^1b})\Lambda ^{p-1}+\cdots$$
where ${^mb}$ are diagonal operators obeying ${^mb}_k=q ({^m}b_{k-n}).$

It follows from Theorem \ref{e} that the Lie algebra $\gamma$ acts on  the moduli space $\mathcal{P}$ of pairs $(K,L)$ of difference operators obeying $KL=qLK$ (we assume that $L$ is monic).
In other words, the moduli space we consider is invariant with respect to the discrete KP-hierarchy.
%To verify this statement we should notice that the flag $V(t)=g(t)V$ where $V$ obeys the conditions of Theorem \ref {e} is compatible with $\Lambda ^q$ and with the operator 
 In the proof of Theorem \ref {e} we use the characterization of difference operators as  pseudodifferential operators compatibble with the filtration $H_m$. We can use instead the characterization of difference operators as pseudodifference operators preserving $H.$    
 
 We obtain
 \begin {thm}
 \label {ee}
 There exists one-to-one correspondence between pairs $(K,L)$ of difference operators obeying $KL=qLK$ and  $L$ is monic of order $n$ and pairs $(\tilde K, V)$ where $V$ is a flag space invariant with respect to $\Lambda^n$ and pseudodifference operator $\tilde K$ obeying $\tilde K\Lambda ^n=q \Lambda ^n \tilde K$

 If the flag space $V$ is invariant also with respect $\Lambda ^{-n}$ then $L^{-1}$ is a difference operator. Conversely, if $L^{-1}$ is a difference operator then $V$ is $\Lambda ^{-n}$-invariant.
 \end{thm}  

If the flag space $V=span (\cdots,w_m,\cdots)$ is invariant with respect to the operators  $\Lambda ^n$  and $\Lambda ^{-n}$ then the vectors $w_0,...,w_{n-1}$ form a $z^{\pm n}$-basis  of $V=\bigcup V_m$ (i.e. the vectors $z^{mn}w_i$ where $m\in\mathbb{Z}, 0\leq i<n$ form a basis of this space). 
To prove this fact we notice that  the operators $\Lambda ^n$ and $\Lambda ^{-n}$  specify a structure of a torsion-free  finitely generated module over the algebra of Laurent polynomials  on $V$. Such a module is always free. ( To prove this one can use the fact that the algebra of Laurent polynomials is isomorphic to the group algebra of $\mathbb{Z}.$) Using the dimension count one can check that the rank of this module is equal to $n$ and  the vectors $w_0,...,w_{n-1}$ are free generators of the module. This is equivalent to the statement we need.

 In the conditions of Theorem \ref {ee}  the operator $\tilde K$  acts on $V=\bigcup V_m$, hence we can consider the matrix $M_i^j$ of $\tilde K$ in the $z^{\pm n}$- basis we constructed (or more generally in any $z^{\pm n}$- basis). By definition this matrix  is the companion matrix of the pair $(K,L).$ (The entries of this matrix are polynomials of $z^n$ and $z^{-n}.$)  Another definition of the companion matrix is based on the remark that  in the case when $L$ is a monic difference operator of order $n$ and $L^{-1}$ is also a difference operator the elements $z^i$ with $0\leq i<n$ constitute a $(L,L^{-1})$-basis of $H$. We can define the companion matrix as matrix of the operator $K$ in this basis (or more generally in any $(L,L^{-1})$-basis.  In this section we always consider the companion matrix in the special basis described above.

It follows from this definition that
\begin{equation}
\label{c}
\sum({^mb})\Lambda ^{p-m}w_i=M_i^jw_j.
\end{equation}
Our goal is to find a pair of difference operators having companion matrix $M$. It will be more convenient  to work with $u_i=z^{-i}w_i$  and with the matrix  $B_i^j(z)=z^{j-i}M^j_i(z^q).$ Then we  should consider the equation
\begin{equation}
\label{cc}
\sum({^mb})\Lambda ^{p-m}(z^iu_i)=z^iB_i^ju_j
\end{equation}
that can be written as 
\begin{equation}
\label{ccc}
\sum_{m\geq 0, k\geq 0} {^mb}_{p-m-k+i}{^ku}_iz^{p-m-k}=B_i^ju_j.
\end{equation}
Here 
$$u_i=\sum_{k\geq 0} {^ku}_iz^{-k}$$
and 
\begin{equation}
\label{bb}
^mb_k=q ({^m}b_{k-n}).
\end{equation}  We assume that the matrix $$B_i^j=\sum _   {m\geq 0} {^mB}^j_iz^{s-m}$$ is known. We should find $^mb_k$ and ${^mu}_i$ by induction. First of all looking at the leading terms we see that $p=s$ and
\begin{equation}
\label{u}
(^0b_{p+i})(^0u_i)=(^0B_i^j)(^0u_j). 
\end{equation}

The recursion formula formula is similar to the formula in Section 2:

\begin{equation}
\label{rr}
(^0b_{p-m+i})(^mu_i)-(^0B_i^j)(^mu_j)=(^mB_i^j)(^0u_j)-(^mb_{p-m+i})(^0u_i)+{\rm  known }\: {\rm terms}.
\end{equation}
To guarantee the existence of solutions we assume that ${^0b_k}$ is an eigenvalue of ${^0B_i^j}$ for all $k$ and that  ${^0B_i^j}$ has $n$ distinct eigenvalues.  As in Section 2  knowing that $\alpha$ is an eigenvalue of ${^0B_i^j}$  and assuming that $p$ and $n$ are coprime  we can say that  
all eigenvalues have the form $\epsilon ^r\alpha$ where $\epsilon ^n=1.$  Combining this statement with equality (\ref {bb}) we obtain that the solution can exist only in the case 
when $q^n=1.$ From the other side if  $q^n=1$  we can take as $^0b_i$ for $0\leq i<n$ arbitrary eigenvalues of ${^0B_i^j}$; then the formula  ${^0b}_k=q({^0}b_{k-n})$ specifies all other $^0b_i$ as eigenvalues of  ${^0B_i^j}. $
We obtain the following
\begin{thm}
\label{tt}
		Let us suppose that the leading term  ${^0B_i^j}$ of the matrix $B_i^j=z^{j-i}M^j_i(z^n)$ has order $p$ that is coprime with $n$. Then it has $n$ distinct eigenvalues . Let us suppose that for every $k$ the number ${^0b}_k$ is equal to one of these eigenvalues. Assume that   ${^0b}_k$ obey  ${^0b}_k=q({^0}b_{k-n})$  and ${^0u_i}$ obey (\ref {u}). Then we can construct a pair of difference operators $K,L$ with  companion matrix $M_i^j(z^n)$  solving the recursion formula (\ref {rr}). The operator $L^{-1}$ also will be a difference operator. 
		\end {thm}

The proof repeats the second proof of Theorem \ref {3}.  Notice that the  arguments used in this proof  give us the numbers ${^mb}_k$ only for $k=p-m+i$ where $0\leq i<n.$ However, knowing these numbers we can find all 	 ${^mb}_k$  from (\ref {bb}).

One can modify the above consideration to study the solutions to the equation  $KL=qLK$  for $q=e^{\hbar}$ in the limit $\hbar\to 0$ as power series with respect to $\hbar.$	
In this situation the equation (\ref{cc}) can be written as 
\begin{equation}
\label{c3}
\sum_{n\geq 0, k\geq 0} {^nb}_{p-n-k+i}{^ku}_iz^{p-n-k}=B_i^ju_j.
\end{equation}
where 
$$u_i=\sum_{k\geq 0, r\geq 0} {^{k,r}u}_iz^{-k}{\hbar}^r,$$
$$^nb_k=\sum_{r\geq 0}{^{n,r}b}_k$$
and 
\begin{equation}
\label{b2}
{^{n,r}b}_k= {^{n,r}b}_{k-q}+\sum _{s>0}{^{n,r-s}}b_{k-q}\frac{1}{s!}.
\end{equation} 
(The last equation follows from (\ref {bb}).) 

We can find   the coefficients $ {^{n,r}u}_i$ and ${^{n,r}b}_k$ using double recursion. We  are writing (\ref {c3}) as a system of equations for these coefficients. The equations for $r=0$ coincide with the equations  coming from (\ref {ccc}) for $q=1$; we have solved them by means of recursion with respect to $n$. Now  we assume that we have found all coefficients with $r<s.$ Then we can find the coefficients  $ {^{n,r}u}_i$ and ${^{n,r}b}_k$ using the recursion formula with respect to $n.$

We obtain 
\begin{thm}
\label{ttt}
		Let us suppose that the leading term  ${^0B_i^j}$ of the matrix $B_i^j=z^{j-i}M^j_i(z^n)$ has $n$ distinct eigenvalues. Then  for every pair $(K,L)$ of commuting difference operators with  companion matrix $M_i^j(z^q)$  we  can find a formal deformation $(K_{\hbar},L_{\hbar})$ having the same companion matrix and obeying $K_{\hbar}L_{\hbar}=e^{\hbar} L_{\hbar}K_{\hbar}.$
\end {thm}
(Saying that $(K_{\hbar},L_{\hbar})$ is a formal deformation of $(K,L)$ we have in mind that $K_{\hbar}$ and $L_{\hbar}$ are power series with respect to  $\hbar$ giving $K,L$ for $\hbar=0.$)

\section {Quantization}
   
Let us consider a pair $(P,Q)$ of commuting differential operators, or a pair $(K,L)$ of commuting difference operators  ( we assume that $Q$ and $L$ are monic and that $L^{-1}$ is also a difference operator). We say that a pair $(P_{\hbar}, Q_{\hbar}) $ of differential operators obeying $[P_{\hbar},Q_{\hbar}]=\hbar$ (quantum curve) is obtained by quantization of the pair $(P,Q)$ if it has the same companion matrix. Similarly, a pair $(K_{\hbar},L_{\hbar})$ of difference operators obeying  $K_{\hbar}L_{\hbar}=e^{\hbar}L_{\hbar}K_{\hbar}$ (discrete quantum curve) is obtained by quantization of the pair $(K,L)$ if it has the same companion matrix. Notice that in these definitions we can work with matrix differential or difference operators. \footnote {Recall that the companion matrix is not unique. We can defined as a matrix of $P$ in the $Q$-basis of $\mathcal{H}_+$ or as a matrix of $K$ in $(L,L^{-1})$-basis of $H$. In the above definition we have in mind the companion matrix in the preferred basis $z^i$ where $0\leq i<q$, where $q$ stands for the order of $Q$ or $L$.}

A  pair $(P,Q)$ of commuting differential operators  satisfies an algebraic equation $A(P,Q)=0.$ This means that $P$ and $Q$ can be considered as meromorphic functions $f,g$ on an algebraic curve $A( x,y)=0.$
Let us describe a procedure \cite {KR} that permits us to construct commuting differential operators starting with two meromorphic functions on an algebraic curve $C$. We start for simplicity with the case when the functions $f,g$ have only one pole at a smooth point $a$ (the function $f$ has a pole of order $p$, the function $g$ has a pole of order $q$). Let us suppose that we have found a subspace $\mathcal{E}$ of the space of meromorphic functions on $C$ having the following properties a) the space   $\mathcal{E}$  contains precisely one function (up to a constant factor) that has a pole of order $n$ at the point $a$ and   these functions span $\mathcal{E}$ (here $n\in \mathbb{N}$ or $n=0$; saying that a function has a pole of order 0 we have in mind that it it is holomorphic and does not vanish at $a$), b) this space is invariant with respect to multiplication by $f$ and $g$. We introduce a coordinate $z$ in the neighborhood of the point $a$ in such a way $g=z^q$ in this neighborhood (hence $z=\infty$ at $a$). Considering the Laurent series (with respect to $z$)  of functions belonging to $\mathcal{E}$ we obtain a subspace $V\subset \mathcal{H}$; it is easy  to check that  we can apply Theorem \ref {2} for $\hbar=0$ to operators of multiplication acting on $V\in Gr$ to get commuting differential operators.

Now we should describe the construction of the space $\mathcal{E}.$ The first idea is to take as  $\mathcal{E}$ the space of all functions having a pole only at $a$.  However, this does not work -the projection of the corresponding subspace of $\mathcal{H}$ on $\mathcal{H}_+$ has the index ${\rm g}(C)-1$ where ${\rm g}(C)$ denotes the genus of the curve $C$  (this follows from well known fact that the sequence of orders of poles  has ${\rm g}(C)$ gaps).
We  should extend this space allowing functions with poles at the points of some divisor (the extended space is still invariant with respect to multiplication by $f$ and $g$).  Generic divisor of degree $g-1$ (non-special divisor) will give the subspace we need ( any divisor of degree ${\rm g}(C)-1$ gives a subspace of index $0$, generically it belongs to the big cell).

A pair $f,g$ of meromorphic functions on the curve $C$ embeds this curve into $\mathbb{C}\times \mathbb{C}$.
If the points of embedded curve satisfy the equation  $A(x,y)=0$ the pair $(P,Q)$ of commuting differential operators we constructed  obeys the same equation. We can quantize this pair  constructing  differential operators obeying the equation $[P_{\hbar},Q_{\hbar}]=\hbar$ and having the same companion matrix in the preferred basis. 
It is important to notice that the explicit construction of the  pair $(P,Q)$ is not necessary in the quantization procedure. We can define the companion matrix directly in terms of functions $f,g.$ Namely, we should find a $g$-basis of the space $\mathcal{E}$ and calculate the matrix of multiplication by $f$ in this basis.

The construction we have described can be generalized to the case when functions $f$ and $g$ have multiple poles. Let us suppose, for example,  that the function $g$ has $s$ poles, all of order $q$, at the  points $a_1,\cdots ,a_s$.  Let us introduce the coordinate $z_i $ in the neighborhood of $a_i$ in such a way that $g=z_i^q$ in this neighborhood (hence $z_i=\infty$ at $a_i$). Then we should find a subspace $\mathcal{E}$ of the space of meromorphic functions on $C$ having the following properties a) for arbitrary integers $n_1,\cdots n_s$ obeying $n_i\geq 0$ the space   $\mathcal{E}$    contains   precisely one function  that behaves as $z_i^{n_i}$ +lower order terms in the neighborhood of $a_i$   and these functions span $\mathcal{E}$, b) this space is invariant with respect to multiplication by $f$ and $g$. To every meromorphic function on $C$ we should assign  $s$ Laurent series representing this function in the neighborhoods of points $a_i$. This construction sends $\mathcal{E}$ to a subspace $V\subset \mathcal{H}^s$; it follows from the condition a) that $V\in Gr_s$. Applying Theorem \ref{g} for $\hbar=0$ we obtain a pair of commuting differential matrix operators. 

To construct the space $\mathcal{E}$ obeying the above conditions we start with with the space of meromorphic functions having poles only at the points $a_1,\cdots,a_s.$ If $f$ has poles only at $a_1,\cdots,a_s$ this space is invariant with respect to multiplication by $f$ and $g$ (property b) ). This property is preserved if we allow additional poles of order $1$ at the points of some divisor. Choosing the divisor appropriately we can satisfy the property a).

Very similar construction works for difference operators. Let us consider an algebraic curve   $C$ and two meromorphic functions $f,g$ that are holomorphic everywhere except two smooth points $a$ and $b$. We introduce a coordinate $z$ in the neighborhood of $a$ in such a way that $g=z^q$ in this neighborhood (we assume that $z=\infty$ at  $a$ hence $g$ has a pole of order $q$). Let us suppose that  a subspace $\mathcal{E}$ of the space of meromorphic functions on $C$ has the following properties a) it is spanned by functions $s_n, n\in \mathbb{Z   }$  such that  $s_n$ has a pole of order $n$ at $a$ for non-negative $n$ and zero of order $-n$ for negative $n$, b) it  is invariant with respect to multiplication by $f$ and $g$.The functions $s_n$ specify a flag in  $\mathcal{H},$  the space $\mathcal{E}$ can be identified with  the corresponding flag space. (To define this flag we consider Laurent series of $s_n$ at the point $a$ with respect to the coordinate $z$). This flag is compatible with multiplication by $f$ and $g.$ Applying Theorem \ref {e} in the case $q=1$ we obtain two commuting difference operators $K$,$L$. If the space $\mathcal{E}$ is invariant also with respect to the multiplication $g^{-1}$ the operator $L^{-1}$ is also a difference operator.

Now we should explain how to find the subspace $\mathcal{E}$ satisfying the conditions a) ,b) . Let us consider the space $\mathcal{K}$ consisting of  meromorphic sections of some line bundle having poles only at points $a,b$ (equivalently we can talk about meromorphic functions on $C$ having poles of any order at $a$ and $b$ , that can also have simple poles at the points of some divisor).  Let us denote by $\mathcal{K}(na)$ the subspace $\mathcal{K}$ consisting of functions of the form $z^{n}f(z)$ where $f$ is finite at $a$; the space $\mathcal{K}(nb)$ is defined in similar way. One can prove that for appropriate choice of $\mathcal{K}$ the space  $\mathcal{K}(na)\cap\mathcal{K}((-n+1)b)$ is one-dimensional; the function $s_n$
can be defined as non-zero element of this space.

 \section {D-modules}

In \cite {D} quantum curves were studied from the viewpoint of $D$-modules. The approach of this paper is closely related to the approach of \cite {SCH} and present paper.  The construction of the point of Grassmannian used in \cite {SCH} plays fundamental role also in \cite {D}. The companion matrix specifies a meromorphic connection 
\begin{equation}
\label{con}
\nabla=\hbar \frac{d}{dz^q}- B^j_i(z),
\end{equation}
the flat sections of this connection can be identified with solutions to the equation (\ref {6}). The flat connection $\nabla$ can be identified with $D$-module studied in \cite {D} (better to speak about the family of connections and a family of $D$-modules parametrized by $\hbar$ or about $D_{\hbar}$-module). 
An equivalent $D$-module can be defined by means of the matrix $M^i_j.$  As we have noticed the matrices $M$ defined by means of different $z^q$-bases are related by a gauge transformation; this means that $M$ can be regarded as a connection and in the definition of $D$-module we can use any $z^q$-basis of $V$ (any $Q$-basis of $\mathcal{H}_+$).

We consider  connections in the neighborhood of $z=\infty$ (connections on the punctured formal disk).
A classification of connections on the punctured formal disk is well known (see \cite {LEV}; we follow \cite {GRA} in the description of the classification).Such a connection can be written in the form
$\nabla=\frac{d}{dz}+A(z)$  
where $A(z)$ stands for the matrix with entries from the field of Laurent series $K=\mathbb{C}((z)).$ (The elements of $K$ are formal series $\sum a_nz^n$ where $a_n=0$ for $n>>0.$)
Notice that $\nabla$ can be characterized as  obeying Leibniz identity  $\mathbb{C}$-linear operator  in finite -dimensional vector space over $K$ . Let us denote this vector space by $V$; the dimension of this space over $K$  is called the dimension of connection. We say that two connections $\nabla$ and $\nabla '$ are gauge equivalent  if there exists an invertible matrix $g$ with entries from $K$ obeying $g\nabla=\nabla 'g.$

Together with the field $K$ we  will consider the field $K_q=\mathbb{C}((v))$ where $v^q=z.$  Notice $K_q$ can be considered as a $q$ -dimensional vector space over $K.$ 
An important example of  connection is  specified by an operator $\nabla_{f,q}=\frac{d}{dz}+\frac{f(v)}{z}$ acting in $K_q$. Here 
$\frac{d}{dz}=\frac{1}{qv^{q-1} }\frac{d}{dv}$ and $f$ stands for an operator of multiplication by Laurent series $f\in K_q.$  The operator 
$\nabla_{f,q}$ determines a $q$-dimensional connection over $K.$ 
Without loss of generality we can assume that 
\begin{equation}
\label{f}
f=\sum_{m\in A} c_mv^m+c_0=\sum_{m\in A} c_mz^\frac{m}{q}+c_0
\end{equation}
where the set $A$ consists of  natural numbers  and $0\leq c_0<1.$ (Connections with other $f$ are gauge equivalent to connections with $f$  in  (\ref {f}).)
All irreducible connections are gauge equivalent to connections of the form (\ref {f}) . However, not all connections of this kind are irreducible; the condition of reducibility is the existence of non-trivial common divisor of the numbers $m\in A$ and $q$ (if such a divisor does exist then we can represent the connection in similar form with $q$ replaced by $q'<q$).

A  more general example of connection is given by an operator 
$$\nabla_{f,q, J}=\frac{d}{dz}+\frac{f(v)}{z}+\frac{J}{z}$$
acting on the space $K_q^n.$  Here $J$ stands for $n\times n$ matrix with complex entries, $f\in K_q.$
 
One can prove that every connection is gauge equivalent to a direct sum of  connections of the form $\nabla_{f,q, J}$ where $J$ is a Jordan block. 

Let us consider a pair $P,Q$ of differential operators obeying $[P,Q]=\hbar$ assuming that $Q$ is normalized and the orders $p,q$ of $P,Q$ are coprime. If $P,Q$ are scalar operators then it follows from our considerations that the corresponding connection is irreducible over the field $K(z^q)$ (over the field $K(z)$ it can be represented as a direct sum of $q$ one-dimensional connections; this statement is equivalent to the diagonalization of the equation (\ref {7}) used in Sec. 2). If $P,Q$ are matrix operators of the size $r\times r$ then the connection is a direct sum of $r$ irreducible connections.
 
It is well known that every $D$-module defined by meromorphic connection in dimension 1 has rank $1$; it can be represented in the form $\mathcal{D}/\mathcal{D}\cdot P$ where $\mathcal{D}$ denotes the algebra of differential operators with meromorphic coefficients (i.e.  polynomials with respect to $\partial _z$ with coefficients that  are  meromorphic with respect to $z)$ and $P\in \mathcal{D}.$ This is the form mostly used in \cite {D}.

The above statement means that the system of equations for flat sections $\nabla u=0$ is equivalent to a single differential equation. One can use this statement to rewrite  the system (\ref {6}) or (\ref {4}) as a single equation
$$\hat A w_0=0,$$
where $\hat A$  is a differential operator with meromorphic coefficients.  Namely, we should consider $w=(w_0,\cdots,w_{q-1}) $ as an element  of $\mathcal{F}^q$  where $\mathcal{F}$ denotes the field  of meromorphic functions. Then $w_0=<e_0,w>$ where $e_0=(1,0,\cdots,0)$ and  $<a,b>=\sum a_ib_i\in \mathcal{F}$. Defining $\nabla_*$ by the formula
$$\nabla_*=\hbar\frac{d}{dz^q}+B_i^j(z),$$
and using $\nabla w=0$ we obtain that
$$(\hbar\frac{d}{dz^q})^s w_0= <\nabla_*^s e_0, w>.$$
To find $\hat A$ we notice that  the vectors $\nabla_*^s e_0$ with $s=0,\cdots q$ are linearly dependent  in $q$-dimensional vector space over $\mathcal{F}$. If
$$\sum_{0\leq s\leq q} a_s(z,\hbar)\nabla_*^s e_0=0$$
we can take 
\begin{equation}
\label{ha}
\hat A=   \sum_{0\leq s\leq q} a_s(z,\hbar)(\hbar\frac{d}{dz^q})^s.
\end{equation}

 Notice that knowing an operator annihilating $w_0$ it is easy to find an operator annihilating $u_0=v_0$:
 $$(\hat A-b(z))u_0=0.$$
 
 It was suggested in the paper \cite {GS} that one should understand a quantum curve as an equation 
 \begin{equation}
\label{ap}
\hat A \Psi =0
\end{equation}
 where 
 $\hat A$ has the form (\ref {ha}) with polynomial coefficients $a_s$. It follows from the above consideration that  starting with a quantum curve in our sense one can construct a quantum curve in the sense of \cite {GS} if one allows meromorphic coefficients $a_s$.
 If $\hat A$ has an irregular singular point at infinity one can construct formal solutions of (\ref {ap}) in the form
 $$\Psi=e^{Q(t)}t^{\rho}\psi(t)$$
where $z=t^r$, $Q(t)$ is a polynomial and $\psi$ is a series with respect to  $t $ and a polynomial with respect to $\log t$. These solutions can be put together in equivalence classes; roughly speaking an equivalent solution  can be obtained by means of substitution $t\to \epsilon t$ where $\epsilon ^r=1$. Instead of $\hat A$ and equation (\ref {ap}) one can consider a connection and corresponding flat sections . The description of the formal solutions of (\ref {ap}) follows then from  the representation of connection as  a direct sum of  connections of the form $\nabla_{f,q, J}$ .( If the Jordan block has size $j\times j$ then the corresponding solutions constitute $j$ classes; each class consists of $q$ elements).
If we $\hat A$ comes from scalar differential operators obeying $[P,Q]=\hbar$ and $Q$ is a monic differential operator of order $q$  then the corresponding connection has the form $\nabla _{f,q},$ hence all solutions are equivalent.

The paper \cite {BAR} is devoted to description of an algorithm that allows us not only to find formal solutions of (\ref {ap}) , but also to find their equivalence classes (rational Newton algorithm). The algorithm is based on the consideration of Newton polygon, equivalent solutions correspond to the same edge of this polygon. If all solutions are equivalent the Newton polygon has only one edge. This gives another proof of the condition satisfied by companion matrix in scalar case. (See the end of Sec 2). From the other side, if the order of the  symmetry group of $\hat A$ is equal to the order of $\hat A$ considered as differential operator and symmetry transformations acting on solutions do not have fixed points , then there exists only one equivalence class.

In particular, the polynomial $\hat A=\hbar^2\frac{d^2}{dz^2}+ s(z)$ is obviously symmetric with respect to the transformation changing the sign of the derivative $\frac{d}{dz}$, hence we can use the above remark to construct a quantum curve in the sense of present paper. Polynomials of this kind were considered in the paper \cite {DM} that is devoted to the generalization and analysis of Eynard-Orantin construction for Hitchin fibration. The authors of this paper show that the quantum curves in the sense of \cite {GS} corresponding to these polynomials appear in the neighborhood of branch points. The above remark shows that this is true also for quantum curves in our sense.

{\bf Acknowledgements} 

I am  indebted to  S. Gukov, M. Luu, M. Mulase, N. Orantin
and V. Vologodsky  for useful  discussions.  I was partially supported
by the NSF grant DMS-0805989.
\bibliographystyle{amsplain}
%\bibliography{sampartb}

\end{document}